\documentclass[conference,compsoc]{IEEEtran}
\pdfoutput=1 

\usepackage{amsmath, amssymb}
\usepackage{algorithm}
\usepackage{algorithmicx}
\usepackage{algcompatible}

\usepackage{graphicx}
\usepackage{tikz}
\usepackage{adjustbox}
\usepackage{float}
\usepackage{subfig}
\usepackage{epsfig}
\usepackage{siunitx}

\usepackage{booktabs}
\usepackage{multirow}
\usepackage{cellspace}
\usepackage{threeparttable}

\usepackage{xspace}
\usepackage{xcolor}
\usepackage{url}
\usepackage{hyperref}
\usepackage{titlesec}
\usepackage{titletoc}
\usepackage{paralist}

\usepackage{cite}

\titlespacing*{\subsubsection}{0pt}{\baselineskip}{\baselineskip}

\graphicspath{{images/}}

{\begin{list}{}%
        {\setlength{\leftmargin}{#1}}%
       \item[]%
}
{\end{list}}

\begin{document}

\title{FaaSGuard: Secure CI/CD for Serverless Applications – An OpenFaaS Case Study\\[-0.8em]}







\author{
\IEEEauthorblockN{Amine Barrak\IEEEauthorrefmark{1}, Emna Ksontini\IEEEauthorrefmark{2}, Ridouane Atike\IEEEauthorrefmark{3}, and Fehmi Jaafar\IEEEauthorrefmark{3}}
\IEEEauthorblockA{\IEEEauthorrefmark{1}\textit{Department of Computer Science and Engineering, Oakland University}, Rochester, MI, USA}
\IEEEauthorblockA{\IEEEauthorrefmark{2}\textit{University of North Carolina Wilmington}, Wilmington, NC, USA}
\IEEEauthorblockA{\IEEEauthorrefmark{3}\textit{University of Quebec at Chicoutimi}, Chicoutimi, QC, Canada \\
Email: aminebarrak@oakland.edu}
}

\maketitle

\newcommand{\emna}[1]{\textcolor{red}{#1}}

\begin{abstract}
Serverless computing significantly alters software development by abstracting infrastructure management and enabling rapid, modular, event-driven deployments. Despite its benefits, the distinct characteristics of serverless functions, such as ephemeral execution and fine-grained scalability, pose unique security challenges, particularly in open-source platforms like OpenFaaS. Existing approaches typically address isolated phases of the DevSecOps lifecycle, lacking an integrated and comprehensive security strategy. To bridge this gap, we propose FaaSGuard, a unified DevSecOps pipeline explicitly designed for open-source serverless environments. FaaSGuard systematically embeds lightweight, fail-closed security checks into every stage of the development lifecycle—planning, coding, building, deployment, and monitoring—effectively addressing threats such as injection attacks, hard-coded secrets, and resource exhaustion. We validate our approach empirically through a case study involving 20 real-world serverless functions from public GitHub repositories. Results indicate that FaaSGuard effectively detects and prevents critical vulnerabilities, demonstrating high precision (95\%) and recall (91\%) without significant disruption to established CI/CD practices.
\end{abstract}

\vspace{0.1cm}

\section{Introduction}
\label{sec:introduction}

Serverless computing, or Function-as-a-Service (FaaS), is reshaping software development by abstracting infrastructure management and enabling event-driven execution~\cite{Hassan2021SurveyServerless}. This lets developers focus on function logic while the runtime handles provisioning, scaling, and fault tolerance. While commercial offerings like AWS Lambda and Azure Functions dominate~\cite{wen2023rise}, open-source platforms such as OpenFaaS, Knative, and Kubeless are gaining traction, allowing deployment on self-managed infrastructure with transparency, extensibility, and flexibility often lacking in commercial cloud platforms.


The shift to serverless computing fundamentally transforms the DevOps lifecycle. Unlike traditional pipelines built for long-running services, serverless applications comprise ephemeral, stateless functions triggered by diverse asynchronous events~\cite{Sharma2019ServerlessPlatform}. This introduces unique delivery characteristics—deep infrastructure abstraction, per-function deployment artifacts, and fine-grained, event-based execution~\cite{Baldini2017ServerlessManifesto}—that disrupt established practices for build automation, CI/CD orchestration, and observability~\cite{Sharma2019ServerlessPlatform}. In open-source settings, developers assume greater responsibility for securing the stack, configuring permissions, and integrating testing and monitoring~\cite{Eskandani2022UphillJourney}, directly influencing how security should be embedded throughout the lifecycle~\cite{marin2022serverless}.

While the security risks of cloud-hosted serverless platforms have been well documented, such as over-privileged functions, dependency vulnerabilities, and injection attacks~\cite{ben2025detection,Shen2019ServerlessSecurity}, the challenges are even more acute in open-source environments. Each stage of the DevOps pipeline presents distinct concerns. During planning, threats related to multi-function interactions and event-based trust boundaries must be accounted for~\cite{marin2022serverless}. In the development phase, insecure code and dependencies may propagate silently due to a lack of standardized static analysis tooling~\cite{wen2023rise}. Testing is frequently hindered by the difficulty of replicating production-grade event flows. In deployment, misconfigured secrets, permissions, or container images can lead to privilege escalation or data leaks~\cite{rahman2023security}. Finally, runtime monitoring is challenged by the short lifespan of function executions and the lack of unified logging across components~\cite{Shen2019ServerlessSecurity}. Unlike monolithic or microservices architectures, serverless systems, particularly in open-source deployments, lack mature, integrated security pipelines~\cite{Ni2024SecurityQuantification}.


To our knowledge, prior work on DevSecOps for serverless platforms has focused mainly on isolated stages of the software delivery lifecycle. Examples include static analysis to detect vulnerable third-party dependencies during development~\cite{marin2022serverless}, event-trigger abuse mitigation during deployment~\cite{Ni2024SecurityQuantification}, information-flow control in event-driven workflows~\cite{datta2020valve}, and anomaly detection at runtime through log-based monitoring~\cite{ben2025detection}. However, full-lifecycle security integration remains understudied, especially for open-source FaaS platforms like OpenFaaS, where practitioners must secure the stack without cloud provider-managed tooling.

In this paper, we address this gap by designing and evaluating a comprehensive DevSecOps pipeline that supports security integration across all phases of the serverless lifecycle. Our approach is tailored specifically for open-source environments and emphasizes practical applicability, architectural transparency, and lifecycle completeness.


The complete replication package is available at \url{https://sites.google.com/view/scam2025}.

\section{Background}
\label{sec:background}
The following taxonomy distills the attack categories most frequently discussed in recent studies that analyse security risks of serverless systems

\begin{itemize}
  \item \textbf{Event Injection (Event Spoofing).}
        Forging or replaying HTTP/queue/cron triggers so a function runs with attacker-controlled
        input~\cite{marin2022serverless, Ni2024SecurityQuantification}.
  \item \textbf{Privilege Escalation.}
        Mis-scoped roles (e.g., wildcard IAM policies or over-broad Kubernetes RBAC)
        allow functions to gain excessive permissions.
  \item \textbf{Code \& Dependency Vulnerabilities.}
        Classic flaws—SQL/command injection, hard-coded secrets, or malicious/typosquatted packages—execute inside the function at runtime\cite{ben2025detection}.
  \item \textbf{Container / Image Misconfiguration.}
        Build pipelines may produce images that run as \texttt{root}, expose extra ports, disable kernel protections, or include outdated libraries. \cite{Liu2022, wong2023security}.
  \item \textbf{Infrastructure-as-Code (IaC) Weaknesses.}
        Insecure Terraform / Helm / Kubernetes YAML (misconfigurations, secret leaks, or injected resources) is applied during deployment\cite{kavas2023architecting}.
  \item \textbf{State Drift \& State-File Compromise.}
        Stealing or tampering with \texttt{terraform.tfstate} hides backdoors or rogue resources from configuration management\cite{usingreverse}.
  \item \textbf{Resource-Exhaustion DoS.}
        Unbounded concurrency or oversized payloads consume CPU/RAM until legitimate functions starve; in public clouds this also drives up cost (``Denial of Wallet'')\cite{kelly2021denial, xiong2021warmonger}.
\end{itemize}

\section{Related Work}
\label{sec:relatedwork}

\subsection{Security Threats in Serverless Platforms}
Serverless computing (Function-as-a-Service, FaaS) enables event-driven applications without direct infrastructure management. Platforms like AWS Lambda and Azure Functions offer built-in isolation, while open-source solutions (e.g., OpenFaaS, Knative) provide configurable controls but introduce distinct security risks\cite{marin2022serverless, barrak2024securing}.

Injection attacks remain prevalent, exploiting unvalidated event data across diverse triggers (HTTP, file uploads, queues), especially in rapid deployments\cite{marin2022serverless}. Side-channel threats also exist; Fang et al.~\cite{Fang2022} demonstrated manipulation of cloud schedulers to facilitate cache-based attacks, while Yelam et al.~\cite{Yelam2021} established covert timing channels across function instances.

Economic vulnerabilities include "denial-of-wallet" or resource exhaustion attacks, where attackers inflate usage costs by repeatedly invoking functions\cite{kelly2021denial}, and "Warmonger" attacks exploiting cold-start churn to degrade performance and increase costs\cite{xiong2021warmonger}.

Supply-chain threats involve malicious images through typosquatting in container registries\cite{Liu2022}. Moreover, workflow-aware frameworks proposed by Datta et al.~\cite{datta2020valve} help enforce security policies via invocation path monitoring and simplified access control.

Observability is limited by short execution durations and constrained runtime visibility, complicating real-time attack detection and forensic analysis. Combining provider logs with behavioral analytics shows promise, but robust real-time monitoring remains challenging\cite{marin2022serverless}.

\subsection{DevOps and Serverless Computing}
DevOps teams increasingly adopt FaaS to benefit from cost-efficient, bursty load management. Early studies by Sokolowski\cite{Sokolowski2021} and Puppala et al.\cite{Puppala2024} highlight faster release cycles and reduced infrastructure maintenance.

To manage growing complexity from numerous interconnected serverless functions, developers adopt declarative Infrastructure-as-Code (IaC) blueprints that automate deployment and orchestration\cite{Casale2020}. Despite automation advancements, practical challenges like local testing, third-party library packaging, and configuration persist, as evidenced by extensive developer queries\cite{wen2021empirical}.

Security remains critical within DevOps workflows. Ni et al.\cite{Ni2024SecurityQuantification} quantified security vulnerabilities across serverless functions, revealing how code flaws, IaC misconfigurations, and orchestration errors can propagate through delivery pipelines.

\begin{figure*}[hbt]
    \centering
    \includegraphics[width=0.95\textwidth]{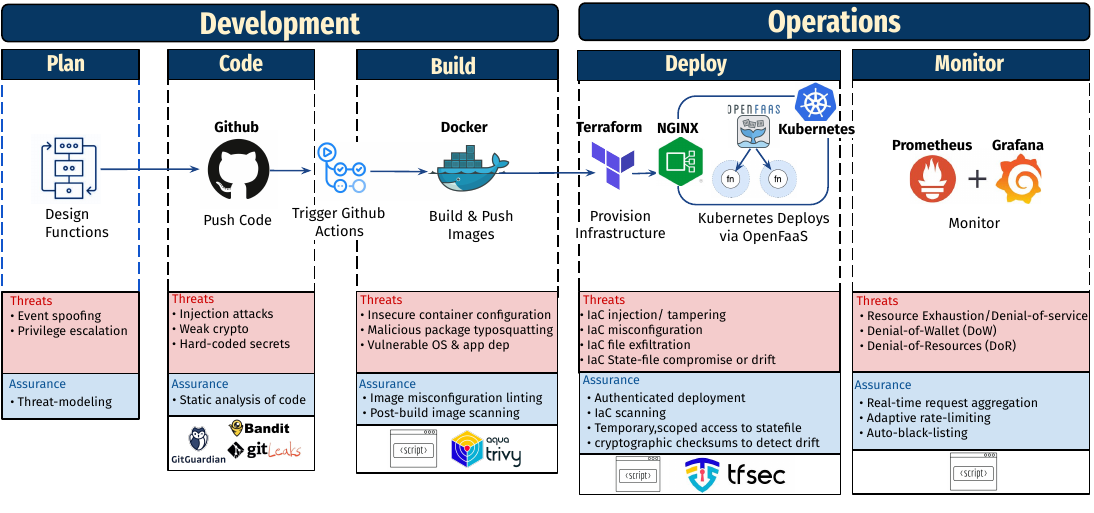}
    \caption{FaaSGard’s DevSecOps pipeline covers Plan, Code, Build, Deploy and Monitor phases, integrating threat modeling, static analysis, signed-image verification and runtime telemetry to secure OpenFaaS functions end to end.}
    \label{fig:types_of_attacks}
\end{figure*}

\section{FaaSGuard Design}  
\subsection{Pipeline Overview}
\label{sec:pipeline-overview}
Figure~\ref{fig:types_of_attacks} illustrates the continuous DevSecOps loop at the heart of \textsc{FaaSGuard}. The loop is divided into two phases that together span the entire serverless life‐cycle.

\underline{\textbf{Development Phase:}}
\begin{itemize}
  \item \textbf{Plan}: Specify functions, event triggers, RBAC bindings, and overall workflows.
  \item \textbf{Code}: Implement the function logic and commit it to the
        version-control repository.
  \item \textbf{Build}: Package source into deployable artefacts
        (e.g., container images).
\end{itemize}

\underline{\textbf{Operation Phase:}}

\begin{itemize}
  \item \textbf{Deploy}: Automatically release the packaged
        functions to the production runtime.
  \item \textbf{Monitor}: Continuously observe performance and behaviour
        through logs, metrics, and tracing data.
\end{itemize}


\subsection{Security objectives per phase}
\label{sec:design-security}
Below we summarise the dominant threat classes at each checkpoint and the
high-level mitigation strategy adopted by \textsc{FaaSGuard}.  Detailed
tooling resides in Section \ref{sec:impl}.

\underline{\textbf{Plan:}}
Specification-level flaws—e.g.\ permissive triggers, over-broad RBAC or
hard-coded secrets—originate here \cite{chowdhuryevaluating}.  FaaSGuard
validates every architectural change against an explicit threat model
before code can be merged.

\underline{\textbf{Code:}}
Source defects such as injection sinks, unsafe crypto and exposed
credentials propagate directly to runtime \cite{ben2025detection}.
A compile-time gate rejects commits whose static-analysis risk score
exceeds a strict threshold.

\underline{\textbf{Build:}}
Supply-chain attacks leverage vulnerable libraries, typosquatted
packages or insecure container settings \cite{Liu2022,
choi2024uncovering, wong2023security}.  FaaSGuard first lint-checks the
build recipe, then scans the resulting image before it reaches a
registry.

\underline{\textbf{Deploy:}}
Infrastructure-as-Code can be tampered with or mis-configured, leading
to privilege escalation or resource exposure \cite{kavas2023architecting,
usingreverse}.  FaaSGuard treats each deployment as an auditable
transaction that must satisfy signed integrity and policy constraints.

\underline{\textbf{Monitor:}}
Runtime attacks—including resource-exhaustion “denial-of-wallet”
scenarios—demand continuous observation \cite{kelly2021denial,
xiong2021warmonger}.  Lightweight anomaly detectors flag deviations and
enforce adaptive rate limits; resulting telemetry feeds future planning.

\section{Implementation on OpenFaaS}
\label{sec:impl}
To validate \textsc{FaaSGuard} with real infrastructure, we realised the entire pipeline on \textbf{OpenFaaS}—the most-starred open-source FaaS platform  (over 24 k GitHub stars) \cite{decker2022performance}. Reference functions are written in \textbf{Python}, the language most prevalent in serverless code and tutorials \cite{spillner2019quantitative}. OpenFaaS’s Kubernetes-native stack separates \emph{build}, \emph{deploy},
and \emph{gateway} layers, allowing each security checkpoint to be instrumented with commodity tooling (GitHub Actions, Docker, Terraform, and an NGINX Ingress Controller).  The resulting workflow—fully reproducible with a single repository—appears in
Figure~\ref{fig:types_of_attacks}.

\underline{\textbf{Plan gate:}}
Every function ships with a \texttt{threat-model.yaml}.  When a pull request modifies this file, the composite \textsc{PlanLint} action executes.  
(i) Pykwalify \footnote{\url{https://github.com/Grokzen/pykwalify}} validates the YAML schema, ensuring mandatory fields and proper typing.
(ii) A repository-local bundle for {Open Policy Agent} (OPA)—a policy-as-code engine that evaluates declarative constraints—verifies architectural rules such as narrowed source-IP filters and vault-backed
secret references. Any violation returns \textsc{FAIL}, blocking the merge and keeping policy and code in lock-step.

\underline{\textbf{Code gate:}}
A push to the protected branch starts the static-analysis stage: \textit{Bandit}\footnote{\url{https://github.com/PyCQA/bandit}} scans the modified Python handlers and tags findings with CWEs, while \textit{Gitleaks}\footnote{\url{https://github.com/gitleaks/gitleaks}} and \textit{GitGuardian \texttt{ggshield}}\footnote{\url{https://github.com/GitGuardian/ggshield}} sweep both application and IaC diffs for embedded secrets. All results are flattened into a protobuf record capturing file path, CWE, severity, confidence, and a remediation hint.
The gate applies a fixed decision table, identical to the default
“break-the-build” policy in GitHub Advanced Security’s CodeQL
workflow \footnote{https://docs.github.com/en/enterprise-server@3.12/code-security/code-scanning/managing-code-scanning-alerts/about-code-scanning-alerts}, which blocks the commit only when a finding is flagged \emph{High} or \emph{Critical} and the accompanying confidence is at least \emph{Medium}.
These two thresholds are read from a YAML settings file so project owners can tighten or relax them without changing code. The scan finishes in a few seconds on typical diffs, keeping feedback tight enough for everyday development.

\underline{\textbf{Build gate:}}
\textit{Trivy}—an all-in-one scanner that inspects container images, file systems, Git repositories, and Kubernetes clusters for CVEs, misconfigurations, and leaked secrets\footnote{\url{https://github.com/aquasecurity/trivy}}—enforces security in two stages.
First, during pre-build, \texttt{trivy config} statically lints the Dockerfile, blocking unpinned base tags, missing non-root \texttt{USER}, or absent \texttt{HEALTHCHECK}.
Second, after the OCI image is built, \texttt{trivy image} (i) matches package versions against Trivy’s composite vulnerability database—aggregating the NVD, GitHub Security Advisories, and major distribution feeds—and (ii) applies CIS Docker-benchmark rules to catch runtime hazards such as extra Linux capabilities or read-write mounts of sensitive paths.
Images with unresolved high-severity findings are discarded; only those that pass both phases are pushed to the private registry, preventing tainted artefacts from ever reaching downstream environments.

\underline{\textbf{Deploy gate:}}
Infrastructure-as-Code manifests written in Terraform are partitioned by environment.  Before any plan is applied, \textit{TFSec}—a static security scanner for Terraform\footnote{\url{https://github.com/aquasecurity/tfsec}}—checks the manifests for privilege escalation, open network paths, and insecure storage. The generated plan is then compared with the last approved state; any cryptographic checksum mismatch aborts the rollout to prevent drift.  During the apply, Terraform operates under a short-lived OIDC token with least privilege.  Once the operation completes, the token expires automatically and the new state snapshot is signed and stored in
an immutable bucket, providing forensic traceability.

\underline{\textbf{Monitor gate:}}
OpenFaaS routes all external traffic through an NGINX Ingress Controller. The controller emits JSON logs (client IP, function, timestamp). A lightweight monitoring agent written in Python streams the logs, groups them into sliding windows keyed by client and function, and computes whether a configurable quota has been breached.  Breaches trigger an Ingress patch that places the offender on a deny-list for a cool-down period; the patch is automatically reverted when the period elapses.  All allow, throttle, and block decisions are exported as Prometheus metrics and visualised in Grafana, enabling operators to tune window size, quota, and block duration without redeploying the sidecar.

\section{Evaluation}

The empirical study is organised along the two macro-phases of
the pipeline. Section~\ref{sec:eval-dev} quantifies how accurately the
\emph{development gates} intercept defects at commit time;
Section~\ref{sec:eval-op} measures the effectiveness of the
\emph{operation gates} during roll-out and under live traffic.
\subsection{Development-phase evaluation}
\label{sec:eval-dev}
\underline{\textbf{Corpus.}}
The benchmark consists of \textbf{20} OpenFaaS repositories
(85\,097 non-blank Python lines, 154 Dockerfiles, 63 Terraform modules),
each frozen at a reproducible commit. All projects meet three inclusion criteria: recent activity \texttt{(> 2 commits in the last year)}, OSI-approved licence, and a minimum footprint of 800LoC.

\underline{\textbf{Execution and logging.}}
Each repository is built twice on an identical GitHub-Actions runner
(2 vCPU, 7 GB RAM): a \emph{baseline} job (unit tests + Docker build) and
a \emph{secured} job that inserts the Code- and Build-gates.
Every alert is logged with path, line, tool, severity, and hint; Trivy’s
CVE database is pre-cached to remove first-run variance.

\underline{\textbf{Ground-truth labelling.}}
Two senior security engineers independently inspect every alert; raw
notes are in the replication package.  Findings are \emph{true positives}
when the surrounding artefact demonstrably violates the referenced
CWE/CVE.  Disagreements (18/165) are resolved by discussion, yielding
Cohen’s $\kappa=0.82$.  A stratified 10 \% sample of functions that
triggered no alerts surfaces four hidden defects, giving a 95 \% confidence interval for the false-negative rate of ± 0.05.

\underline{\textbf{Metrics.}}
We report per-gate precision, recall, and $F_{1}$; execution-time overhead is the wall
clock difference between the baseline and secured jobs on the same runner.

Table~\ref{tab:rq1-accuracy} summarises these results.  Of the 165
pipeline alerts, 157 are valid, for overall precision 0.95; the four
hidden defects imply recall 0.91, yielding a macro-averaged
$F_{1}=0.93$.

\begin{table}[h!]
  \centering
\caption{Detection accuracy of FaaSGuard on the twenty-project corpus. 
Recall is extrapolated from the clean-sample inspection (95\% CI $\pm$0.05 false-negative rate).}
  \label{tab:rq1-accuracy}
  \begin{tabular}{lcccccc}
    \toprule
    \textbf{Gate} & \textbf{True} & \textbf{False} &
      \textbf{Precision} & \textbf{Recall} & \textbf{F{1}}\\[-1pt]
                   & \textbf{positives} & \textbf{positives} & & &\\
    \midrule
    Code  & 45  & 2 & 0.96 & 0.88 & 0.92\\
    Build  & 112 & 6 & 0.95 & 0.93 & 0.94\\
    \midrule
    \textbf{Overall}         & \textbf{157} & \textbf{8} &
      \textbf{0.95} & \textbf{0.91} & \textbf{0.93}\\
    \bottomrule
  \end{tabular}
\end{table}

All four false negatives arise from unconventional version tags rather than gaps in the detection rules. Each affected project installs dependencies directly from Git commits, so \textsf{pip} records a placeholder such as \texttt{0.0.0+<sha>} instead of a semantic tag like \texttt{Flask 2.3.1}. Because Trivy correlates CVEs on the pair ⟨package, version⟩, the string \texttt{Flask 0.0.0} lies outside the range 2.3.0–2.3.2 and is ignored. Manually replacing \texttt{0.0.0} with \texttt{2.3.1} in the SBOM prompts Trivy to issue the expected alert, demonstrating that the miss is solely a version-metadata artefact, not a deficiency in the detection logic.


Table~\ref{tab:rq1-categories} contains the 157 validated findings from
the Code- and Build-gate inspection into three threat classes that align
with our extended STRIDE taxonomy. 
Column 2 (\emph{True positives}) counts individual defects;  Column 3 (\emph{Projects affected}) shows how many of the 20 repositories contain at least one instance; 
Column 4 converts that count into a corpus percentage;  
Column 5 reports the \emph{median} CVSS v3.1 base score for the class. We prefer the median to the arithmetic mean because the CVSS distribution is strongly skewed—several 9.8-Critical entries coexist with a handful of 4.0-Medium ones—so the median offers a more robust notion of “typical” severity.

\begin{table}[t]
  \centering
  \caption{Confirmed defects grouped by category and mapped to the STRIDE taxonomy embedded in \textit{FaaSGuard}. CVSS values refer to the median within each category.}
  \label{tab:rq1-categories}

  \setlength{\tabcolsep}{4pt}

  \begin{tabular}{|p{3.3cm}|p{1.1cm}|p{1cm}|p{0.8cm}|p{0.8cm}|}
    \hline
    \textbf{Category / STRIDE mapping} &
    \textbf{True positives} &
    \textbf{Projects affected} &
    \textbf{Corpus share} &
    \textbf{Median CVSS} \\

    \hline
    Vulnerable dependencies (Tampering) &
    103 & 10 & 50\,\% & 8.1 \\
    Code misconfigurations (Injection, Spoofing, Elev.\,Priv.) &
    34 & 4 & 20\,\% & 6.4 \\
    Plain-text secrets (Information Disclosure) &
    20 & 2 & 10\,\% & \textbf{9.8} \\
    \hline
    \textbf{Any defect} &
    \textbf{157} & \textbf{15} & \textbf{75\,\%} & — \\
    \hline
  \end{tabular}
\end{table}

The results reveal three clear patterns.  
First, vulnerable third-party packages (STRIDE Tampering) dominate:
103 distinct CVEs appear in half the projects, and their median CVSS of
8.1 places the typical instance firmly in the High bracket.  
Second, code-level misconfigurations flagged by Bandit (Injection,
Spoofing, Elevation of Privilege) occur in four repositories and carry a
median of 6.4—numerically smaller, yet spanning multiple STRIDE threats,
underscoring the need for static analysis even when dependency scanning
is in place.  
Third, plain-text secret leaks are the rarest phenomenon (two projects)
but the gravest: their median CVSS of 9.8 reflects the fact that a
leaked token can be exploited instantly.
Overall, 15 of the 20 repositories(75 \% of the corpus) contain at least one confirmed defect.

To make the abstract categories concrete, we highlight one
example per class; each is chosen because its issue is
both severe and easy to understand.

\begin{itemize}
\item \textbf{Vulnerable dependencies (Tampering).}  
      \emph{sample-python-app}\footnote{\url{https://github.com/ridgit/create_secure_image/tree/main/functions/sample_python_app_master}}
      ships \texttt{Flask 2.3.1}\,(CVE-2023-30861, cookie disclosure) and
      \texttt{setuptools 69.0.3}\,(CVE-2024-6345, remote code execution).
      Trivy rates both CVEs at CVSS~$\ge 8.8$, illustrating why outdated
      libraries dominate the “High’’ bucket.

\item \textbf{Code misconfiguration (Injection / Spoofing).}  
      \emph{microservicepythontask}\footnote{\url{https://github.com/ridgit/create_secure_image/tree/main/functions/microservicepythontask}}
      is flagged by Bandit B324 for hashing user data with
      \texttt{MD5} and \emph{without} the
      \texttt{usedforsecurity=False} guard.  Because MD5 is
      collision-prone, an adversary can craft inputs with identical
      digests, subverting integrity or password checks.

\item \textbf{Plain-text secrets (Information Disclosure).}  
      \emph{movieservice}\footnote{\url{https://github.com/ridgit/create_secure_image/tree/main/functions/movieservice}}
      commits a production API key in \texttt{settings.py}.  Gitleaks
      classifies the leak as Critical (CVSS 9.8) because the credential
      can be exploited immediately, explaining the high median score for
      this class in Table~\ref{tab:rq1-categories}.
\end{itemize}

\subsection{Operation-phase evaluation}
\label{sec:eval-op}

The runtime component of \textsc{FaaSGuard} must demonstrate that it
(1) blocks unsafe infrastructure changes at deploy time and
(2) mitigates misuse once a service is live. To that end, we deploy
real applications to fresh Kubernetes clusters, inject controlled
faults, and apply workloads that combine benign traffic with
adversarial bursts. All scripts, manifests, and logs are released
in the replication package.

\underline{\textbf{Application set.}}
From the twenty-project corpus we select three services that span the
operational spectrum: a single-function endpoint, a three-function
micro-service, and a five-function data pipeline.  These represent the
smallest, median, and largest artefacts in function count, trigger
diversity, and container footprint.  Each service is deployed to a clean
Kubernetes 1.29 cluster consisting of four OpenStack VMs (8 vCPU,
16 GB RAM, 200 GB SSD) via the reference Terraform recipes.  The
OpenFaaS gateway (v0.28.2) is fronted by an NGINX Ingress Controller
(v1.11); aside from the ingress IP, no node is Internet-routable.

\underline{\textbf{Infrastructure faults.}}
Per service we execute four Terraform plans: one clean baseline and
three plans each embedding a canonical IaC defect—(i) wildcard RBAC on
the gateway service account, (ii) public \texttt{LoadBalancer} without
CIDR restriction, and (iii) a tampered state-file checksum.  The
unmodified CI workflow runs \textit{tfsec}, OPA, and state-integrity
checks exactly as described in §\ref{sec:impl}.  As summarised in
Table~\ref{tab:deploy-summary}, the Deploy gate blocks every injected
fault and accepts every baseline, with a median decision latency of
7.4 s—negligible relative to image pull and pod start-up times.

\begin{table}[h!]
  \centering
  \caption{Verdicts and latency for the Deploy gate
            (median of three repetitions per plan).}
  \label{tab:deploy-summary}
  \begin{tabular}{lcc}
    \toprule
    \textbf{Plan} & \textbf{Verdict} & \textbf{Latency (s)}\\
    \midrule
    Clean baseline  & accept & 7.1 \\
    RBAC wildcard   & reject & 7.5 \\
    Public trigger  & reject & 7.6 \\
    State drift     & reject & 7.6 \\
    \bottomrule
  \end{tabular}
\end{table}

\underline{\textbf{Traffic workloads.}}
Steady traffic is generated with \texttt{k6}\footnote{\url{https://k6.io}}
(version 0.49), and bursts with \texttt{vegeta}\footnote{\url{https://github.com/tsenart/vegeta}}
(version 12.10).  Each cluster is subjected to  
(i) a five-minute steady run at its 95-percentile historic rate,  
(ii) a 30-second flash flood at ten times that rate, and  
(iii) a 15-minute \emph{resource-exhaustion} stream at 1.5 × baseline.  
The Monitor gate records detection latency, malicious-through rate,
false-block rate, and CPU savings relative to an unprotected control.
Aggregate results appear in Table~\ref{tab:monitor-summary}.

\begin{table}[h]
  \centering
  \caption{Performance of the Monitor gate
           (mean of nine runs: three workloads × three services).}
  \label{tab:monitor-summary}
  \begin{tabular}{
    p{1.3cm}  
    p{1.5cm}  
    p{1.6cm}  
    p{1.cm}  
    p{1.2cm}  
  }
    \toprule
    \textbf{Workload} &
    \textbf{Detection latency (ms)} &
    \textbf{Malicious through (\%)} &
    \textbf{Legit lost (\%)} &
    \textbf{CPU cost saved (\%)}\\
    \midrule
    Flash flood      & 180 & 1.8 & 0.6 & 96 \\
    Resource Exhaustion & 240 & 3.2 & 0.4 & 91 \\
    \bottomrule
  \end{tabular}
\end{table}

The Deploy gate neutralises all three injected IaC faults across
heterogeneous services and adds less than eight seconds of roll-out
latency, well within ordinary container-start variance.  The Monitor gate
quashes both burst traffic and slow-drip resource-exhaustion attacks
within 250 ms, letting fewer than four per cent of malicious requests
through while blocking under one per cent of legitimate traffic; the
resulting CPU savings exceed ninety per cent during cost-inflation
scenarios.  All results were obtained with the stock policy bundle and
default parameters shipped with \textsc{FaaSGuard}; no project-specific
tuning was required.  These findings show that the operation-phase
controls generalise beyond the design corpus and impose no measurable
penalty on availability or deployment velocity.

\section{Conclusion}
\label{sec:conclusion}
This paper introduced FaaSGuard, a comprehensive DevSecOps pipeline tailored explicitly to enhance security across the full lifecycle of serverless computing in open-source environments. Our empirical validation using OpenFaaS highlights FaaSGuard's capability to reliably identify and block critical security issues, including dependency vulnerabilities, insecure coding practices, injection attacks, privilege escalation, hard-coded secrets, and resource-exhaustion threats, while ensuring minimal overhead and maintaining compatibility with modern CI/CD processes.

Our evaluation revealed key limitations. The implementation focuses on Python-based functions, limiting generalizability. Runtime threshold calibration is challenging and requires tuning. Third-party tool reliance introduces trust boundaries needing ongoing management.

\bibliographystyle{IEEEtran}
\bibliography{references}
\end{document}